\documentclass[%
 reprint,
superscriptaddress,
 amsmath,amssymb,
prx,
]{revtex4-2}

\usepackage{graphicx}% Include figure files
\usepackage{dcolumn}% Align table columns on decimal point
\usepackage{bm}% bold math
\usepackage{siunitx}
\usepackage{hyperref}% add hypertext capabilities
\usepackage{booktabs} % for \toprule, \midrule, \bottomrule macros

\graphicspath {{figures/}}
\begin{document}

\preprint{APS/123-QED}

\title{Scalable Multispecies Ion Transport in a Grid-Based Surface-Electrode Trap}% Force line breaks with \\

\author{Robert D. Delaney}
\thanks{These authors contributed equally to this work.}
\affiliation{Quantinuum, 303 S. Technology Ct., Broomfield, CO 80021, USA}
\author{Lucas R. Sletten}%
\thanks{These authors contributed equally to this work.}
\affiliation{Quantinuum, 1985 Douglas Dr. N., Golden Valley, MN 55422, USA}
\author{Matthew J. Cich}
\affiliation{Quantinuum, 1985 Douglas Dr. N., Golden Valley, MN 55422, USA}
\author{Brian Estey}
\affiliation{Quantinuum, 303 S. Technology Ct., Broomfield, CO 80021, USA}
\author{Maya I. Fabrikant}
\affiliation{Quantinuum, 303 S. Technology Ct., Broomfield, CO 80021, USA}
\author{David Hayes}%
\affiliation{Quantinuum, 303 S. Technology Ct., Broomfield, CO 80021, USA}
\author{Ian M. Hoffman}
\affiliation{Quantinuum, 303 S. Technology Ct., Broomfield, CO 80021, USA}
\author{James Hostetter}
\affiliation{Quantinuum, 1985 Douglas Dr. N., Golden Valley, MN 55422, USA}
\author{Christopher Langer}
\affiliation{Quantinuum, 303 S. Technology Ct., Broomfield, CO 80021, USA}
\author{Steven A. Moses}
\affiliation{Quantinuum, 303 S. Technology Ct., Broomfield, CO 80021, USA}
\author{Abigail R. Perry}
\affiliation{Quantinuum, 303 S. Technology Ct., Broomfield, CO 80021, USA}
\author{Timothy A. Peterson}
\affiliation{Quantinuum, 1985 Douglas Dr. N., Golden Valley, MN 55422, USA}
\author{Andrew Schaffer}
\affiliation{Quantinuum, 1985 Douglas Dr. N., Golden Valley, MN 55422, USA}
\author{Curtis Volin}
\affiliation{Quantinuum, 303 S. Technology Ct., Broomfield, CO 80021, USA}
\author{Grahame Vittorini}
\email{Grahame.Vittorini@quantinuum.com}
\affiliation{Quantinuum, 1985 Douglas Dr. N., Golden Valley, MN 55422, USA}
\author{William Cody Burton}
\affiliation{Quantinuum, 303 S. Technology Ct., Broomfield, CO 80021, USA}

\date{\today}% It is always \today:day,
\begin{abstract}
 
%To date, Linear arrays of trapped ions have achieved exceptional performance on rigorous quantum computing benchmarks such as quantum volume.  However, a fully two-dimensional quantum charge-coupled device (qccd) trapped ion quantum computer has yet to be realized. 

Quantum processors based on linear arrays of trapped ions have achieved exceptional performance, but scaling to large qubit numbers requires realizing two-dimensional ion arrays as envisioned in the quantum charge-coupled device (QCCD) architecture. Here we present a scalable method for the control of ion crystals in a grid-based surface-electrode Paul trap and characterize it in the context of transport operations that sort and reorder multispecies crystals. By combining cowiring of control electrodes at translationally symmetric locations in each grid site with the sitewise ability to exchange the voltages applied to two special electrodes gated by a binary input, site dependent operations can be achieved using only a fixed number of analog voltage signals and a single digital input per site. In two separate experimental systems containing nominally identical grid traps, one using $^{171}\mathrm{Yb}^{+}$-$^{138}\mathrm{Ba}^{+}$ crystals and the other $^{137}\mathrm{Ba}^{+}$-$^{88}\mathrm{Sr}^{+}$, we demonstrate this method by characterizing the conditional intrasite crystal reorder and the conditional exchange of ions between adjacent sites on the grid. Averaged across a multisite region of interest, we measure subquanta motional excitation in the axial in-phase and out-of-phase modes of the crystals following these operations at exchange rates of 2.5 kHz.  In this initial demonstration, the logic controlling the voltage exchange occurs in software, but the applied signals mimic a proposed hardware implementation using crossover switches.  These techniques can be further extended to implement other conditional operations in the QCCD architecture such as gates, initialization and measurement.  
\end{abstract}

\maketitle
%\tableofcontents
\section{Introduction}
As one of the original platforms suggested for quantum computing more than 25 years ago \cite{cirac1995quantum}, trapped ion systems have maintained their status as a leading candidate for gate-based quantum processors.  Trapped ion qubits have been used to demonstrate the published world records in two-qubit gate fidelity \cite{srinivas2021high,clark2021high}, single-qubit gate fidelity \cite{harty2014high}, and state preparation and measurement error \cite{an2022high}.  In conjunction with the all-to-all qubit connectivity enabled by ion transport operations \cite{wineland1998experimental,kielpinski2002architecture,pino2020demonstration}, trapped ion quantum computers have raised quantum computing performance benchmarks such as quantum volume \cite{cross2019validating} to new heights \cite{moses2023race}.      

Existing trapped ion quantum computers typically have achieved all-to-all connectivity using linear Paul traps, which use a combination of low-bandwidth control electrodes and an rf potential to provide confinement \cite{paul1990electromagnetic, pino2020demonstration, kielpinski2002architecture}.  In the quantum charge-coupled device (QCCD) architecture, the control electrodes are leveraged to create an array of discrete confining locations (“wells”) within the linear trap. Ion pairs cotrapped in a single well are entangled through their shared motional modes \cite{sorensen2000Entanglement}. Arbitrary qubit connectivity is then realized by cotrapping target qubits using operations that merge and separate wells along the trap, interleaved with swap operations that reorder qubits within a well \cite{pino2020demonstration, moses2023race}.

Trapped ion quantum processors with one-dimensional geometry suffer scaling limitations as the number of ions increases.  The increased length of the trap becomes an engineering challenge, while the poor linear scaling of sorting time with qubit count presents a fundamental barrier \cite{malinowski2023wire}.    Some measure of success has been achieved by obviating transport and instead using the collective normal modes of a long linear chain of ions trapped in a single potential well as a bus for interactions among the ions \cite{lin2009large, monz201114,ulm2013observation,debnath2016demonstration, chen2023benchmarking}. Long chains also encounter severe scaling obstacles, including low axial frequencies that are challenging to cool, increased crowding of radial motional mode frequencies, and gate times that scale with chain length \cite{katz2023programmable}.

A proposed solution for scaling the size of ion traps is a 2D (or 3D) grid of short linear traps connected by junctions. Proof-of-concept demonstrations of transport through such junctions have been performed \cite{hensinger2006t,  amini2010toward,moehring2011design, blakestad2011near, wright2013reliable, shu2014heating, decaroli2021design, burton2023transport}, including high-fidelity, low heating transport for  single ions \cite{blakestad2011near} and multispecies crystals  \cite{burton2023transport}.  In particular, multispecies junction transport eliminates the need to separate a qubit ion from its sympathetic coolant \cite{barrett2003sympathetic}, vastly simplifying the transport operations required for ion sorting \cite{burton2023transport} and midcircuit cooling.  

Without a mitigating strategy, the number of control electrodes and associated wiring elements scales linearly with the number of ions in the trap. This “wiring problem” is a common issue that is shared by many quantum computing platforms \cite{krinner2019engineering}. In the ion case, this scaling challenge has been partially alleviated through co-wiring, or “broadcasting,” of individual control signals across multiple surface electrodes in order to perform the same operations on multiple wells in parallel \cite{maunz2016high, moses2023race}. This approach allows storage and sorting of more ions with fewer analog signals but does not fundamentally change the scaling with system size, as individual control electrodes must be incorporated throughout the trapping array in order to perform arbitrary sorting. Recent proposals have suggested \cite{malinowski2023wire, deen2024ion} adding an array of switches in order to dynamically change the mapping between control signals and electrodes based on binary inputs, effectively making the cowiring reconfigurable at each sorting step to achieve both scalability and sitewise control. 

In this work, we experimentally demonstrate the transport operations for sorting qubits in a grid-based rf Paul trap based on a new, scalability-focused transport primitive termed `center to left or right' (C2LR). This primitive enables independent control of the well trajectory in each grid leg using a single digital control signal per site and a fixed number of analog signals independent of grid size. The C2LR primitive can be combined with global shifts and rotations to implement higher-level operations required for a full qubit-sorting algorithm, namely site dependent swapping of qubit pairs between adjacent legs through their shared junction and site dependent reordering of ion crystals. In two experimental setups using nominally identical traps, we test these site-swapping operations with either $^{171}\mathrm{Yb}^{+}$-$^{138}\mathrm{Ba}^{+}$ (Yb-Ba) or $^{137}\mathrm{Ba}^{+}$-$^{88}\mathrm{Sr}^{+}$ (Ba-Sr) two-ion crystals and show that these crystals remain in their motional ground state at swap rates exceeding $\Gamma_\mathrm{swap} = 2.5$~kHz, paving the way for large scale, periodic, 2D ion traps with practically realizable wiring demands.

\section{Experimental Systems}

\begin{figure}%[!h]
\includegraphics[width=\columnwidth]{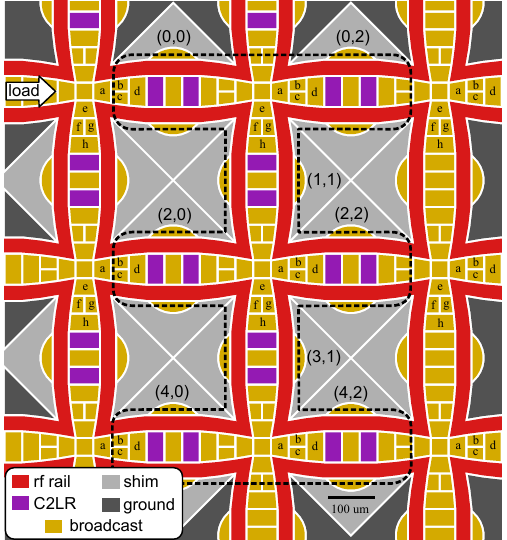}
\caption{\label{fig:fig1} Representative drawing of the relevant trap top metal detailing the rf electrode (red), generic control electrodes (yellow), special center to left or right electrodes (purple), shim electrodes (light gray), and ground plane (dark gray). The rf electrode and ground plane are continuous; the apparently isolated top metal features are connected through traces routed beneath the surface \cite{moses2023race}. The generic control electrodes are cowired between equivalent locations in each leg; lettering on several such electrodes illustrates the pattern. Ions are loaded approximately $\sim$1~mm from the left edge of the depicted region and shifted to one of the enumerated grid sites for testing. The transport characterization that follows is measured at the enumerated grid sites with operations traversing the outlined segments of the grid (dashed line).  Additional tiles of the grid and the geometry of the terminating boundary of the rf electrode located outside of this image serve to increase the uniformity of the potential in the central zones where transport operations are tested.}
\end{figure}

The surface trap presented here is a 2D tiling of square cells, each made of a pair of \SI{375}{\micro\metre} long linear Paul segments ("legs") connected at 90 degrees by an X-junction (Fig.~\ref{fig:fig1}). Two systems use this trap to both demonstrate the flexibility of the design but also to pursue different potential benefits:  Yb-Ba is a thoroughly tested pair with existing demonstrations in junction transport \cite{burton2023transport} as well as commercial operation \cite{pino2020demonstration,moses2023race}, while Ba-Sr is an attractive alternative species combination requiring no ultraviolet light for trapping and control, a  favorable trait for optical systems, in general, and integrated photonic structures in particular \cite{Sorace2019versatile}. Individual ions are loaded through a hole through the entire trap stack using neutral flux from atomic ovens and standard photoionization techniques \cite{olmschenk2007manipulation}. Crystals are formed by shifting ions away from the load hole where a merge operation joins two wells into one.

\begin{figure*}%[!h]
\includegraphics[width =\textwidth]{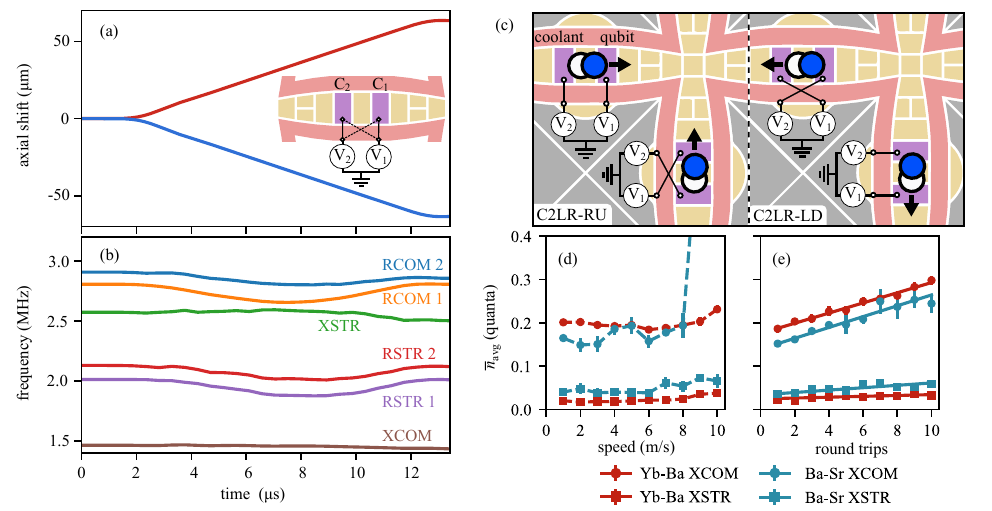}% Here is how to import EPS art
\caption{\label{fig:fig2}  Illustration and validation of the center to left or right primitive. (a) Voltages $V_1$ and $V_2$ can be applied as either $V_1 \rightarrow C_1$, $V_2 \rightarrow C_2$ for a rightward shift (C2LR-R) or $V_1 \rightarrow C_2$, $V_2 \rightarrow C_1$ for C2LR-L.  The red and blue traces are the resulting simulated ion trajectories for the as-solved waveform (C2LR-R) and with C1 and C2 voltages exchanged (C2LR-L). (b) Simulated motional modes of a Yb-Ba crystal during the C2LR waveform, showing: the axial in-phase (XCOM) and out-of-phase (XSTR) modes, two pairs of radial stretch (RSTR 1 and RSTR 2) and radial center-of-mass (RCOM 1 and RCOM 2) modes.  (c) Depiction of C2LR running in two vertical and two horizontal legs. The voltage exchange in the experiment is equivalent to these crossover switches but is realized in software.  (d) Thermal occupancy after cooling and C2LR operations for a single Yb-Ba or Ba-Sr crystal, averaged over both directions and all measurement locations, remains low to speeds of at least 8~m/s.  (e) Thermal occupancy as a function of the number of round-trips at a fixed transport speed of 6 m/s with linear fits, reported in Table~\ref{TAB:TAB_A}.  In (d) and (e), error bars represent one standard deviation on the mean.   }
\end{figure*}
The rf confinement is provided by a single, continuous electrode that uses subsurface routing to connect otherwise separate top metal traces (Fig.~\ref{fig:fig1}).  An rf voltage $V(t) = V_{\mathrm{rf}} \sin\left(\omega_{\mathrm{rf}} t\right)$ with amplitude of $V_\mathrm{rf} \approx 180~V$ is applied to trap Yb-Ba crystals, while a smaller voltage of $V_\mathrm{rf} \approx 167$~V is used to trap Ba-Sr crystals, due to the smaller total mass.  In both experimental setups the rf voltage is driven at a frequency of $\omega_\mathrm{rf} / 2\pi \approx 42$~MHz through a cryogenic rf resonator. Segmented control electrodes located on or near the trap axis provide the dynamic control over the electric potential required for transport operations. In addition, large shim electrodes, located off the trap axis, are used to compensate large-scale, static potential inhomogeneities. The shim electrodes are held at 0~V except during split and merge operations or while the ions are idling in the leg centers. Shimming voltages may be necessary for sorting in larger grids or to meet gating requirements. These voltages, while necessarily site dependent, can be provided by low-bandwidth sources and multiplexed using switches and a sample-hold technique to greatly reduce their cost \cite{malinowski2023wire}.

%In principle these shim electrodes must be wired entirely independently, and are wired independently in this work, but for scaling to larger system size it should be possible to use switch multiplexing and sample/hold techniques to independently control the slowly changing shim voltages \cite{malinowski2023wire}.

%The shim electrodes are required to cancel out stray fields that can emerge at spatially distinct locations in the trap \cite{doret2012controlling}, and are mainly required during split-merge operations, gates and during cooling.  

The majority of the electrodes in the grid are physically cowired according to the translational symmetry of the trap (see the lettering scheme of electrodes in Fig.~\ref{fig:fig1}). In each leg, there are 12 broadcast control electrodes, two shim electrodes and two center to left or right electrodes (C2LR, described in the next section). The junction center electrode included in this count is shared by horizontal and vertical legs, but otherwise all vertical and horizontal electrodes have separate physical wiring.  The electrodes directly adjacent to each junction-center electrode in Fig.~\ref{fig:fig1} are independently wired for testing purposes but are counted among the broadcast electrodes as their voltages are tied together in software for all presented operations. The C2LR electrodes are located on either side of the leg center electrode and can be controlled independently.% While the trap electrodes and rf potential are symmetric under 90 degree rotation, vertical and horizontal legs are wired independently. 

We engineer a highly translationally symmetric rf pseudopotential in the zones highlighted in Fig.~\ref{fig:fig1} by both extending the periodic geometry outward beyond the region of interest and by optimizing the shape of the terminating boundary of the rf electrode. We focus our measurements on the central six horizontal and two vertical legs (dashed line in Fig.~\ref{fig:fig1}) of the 20-junction grid where numerical simulations indicate the deviation in height of the rf potential minimum from the mean varies by less than 0.5\%.

For all transport operations we generate a numerical electrostatic model of the entire trap top metal and use a least-squares optimization procedure to solve for a transport waveform (see Appendix ~\ref{sec:app:waveforms} for more details) in a single unit tile of the trap, namely the tile containing legs (2,0) and (1,1). We then verify in simulation that the waveform broadcasts with sufficient performance across all unit tiles of the central portion of the grid shown in Fig.~\ref{fig:fig1} before testing on the experiment.  %Typically we solve for potential wells in zones (2,0) and/or (1,1) (see labels in Fig.~\ref{fig:fig1}) with symmetric electrodes tied together across unit tiles of the trap.  %We then validate that these voltage waveforms broadcast throughout the central region of the grid.    

For cooling, split or merge and measurement operations the ions are held at the center of a leg at a height of \SI{53}{\micro\metre}, though this height varies during transport by up to several microns.  

\section{Center-to-left-or-right primitive}
At the core of our two-dimensional ion transport protocol is the C2LR operation, where crystals are shifted left or right depending on how two voltages, $V_1$ and $V_2$, are mapped onto the two C2LR electrodes, $C_1$ and $C_2$ (inset Fig.~\ref{fig:fig2}(a)). If the mapping between electrodes and voltage sources is fixed, as is the case in the experimental systems here, then C2LR in conjunction with broadcasting reduces the number of additional analog signals per zone required to control the trap down to two. However, a crossover switch (double pole, single throw) in each leg could be used to dynamically control the mapping between the two voltages and C2LR electrodes with a single binary input. In this case, transport in an $N$-site trap could be controlled with a fixed number of analog signals (here, 27), $N$ digital inputs, and $N$ crossover switches. 

To generate the C2LR primitive, we impose mirror symmetry about the center of the legs on the applied voltages, tying pairs of electrodes on opposite sides of the center together in software. Crucially, the voltages $V_1$ and $V_2$, applied to $C_1$ and $C_2$ respectively, are permitted to vary independently (inset Fig.~\ref{fig:fig2}(a)).  With this single-exception mirror-symmetry constraint, we solve for a linear transport segment \cite{burton2023transport} that shifts the ions 62.5~{\textmu}m to the right of the zone center (Fig.~\ref{fig:fig2}(a) and (c)). Shifting by this 1/6 of the leg length maximizes the spacing between neighboring wells in subsequent transport operations such that crystals are never closer than 1/3 of the leg (\SI{125}{\micro\metre}), enough distance to neglect interaction between ions in separate wells (Appendix ~\ref{CoulombInteraction}) in a densely loaded trap. The wells are constrained such that the axial center-of-mass frequency remains constant during the transport, at $\omega_\mathrm{XCOM}/2\pi = 1.4$~MHz for Yb-Ba and $\omega_\mathrm{XCOM}/2\pi =  1.85$~MHz for Ba-Sr, while the radial mode frequencies are free to vary (Fig.~\ref{fig:fig2}(b)).    
\begin{table}
\centering
$\begin{array}{@{} S[table-format=3.0] ccc @{}} 
\toprule
\textrm{Crystal} & \textrm{XCOM} \ \Delta \bar{n}_\textrm{avg} & \textrm{XSTR} \ \Delta \bar{n}_\textrm{avg}\\
& \quad (\textrm{mq/round~trip)} \quad \quad  & (\textrm{mq/round~trip}) \\
\midrule

\textrm{Yb-Ba}  & 11.7\pm0.5  & 1.0\pm0.3   \\
\textrm{Ba-Sr} & 12.5\pm0.7   & 2.8\pm0.3 \\
\bottomrule
\end{array}$
\caption{\label{TAB:TAB_A} Motional excitation per C2LR round-trip based on sideband asymmetry for Yb-Ba and Ba-Sr crystals at a fixed linear transport speed of 6~m/s (see fit in Fig.~\ref{fig:fig2}(e)), where mq is one thousandth of a quanta} 
\end{table}

The axial symmetry of the electrode geometry about the leg center in conjunction with the constraints described above implies that exchanging the voltages applied to the the C2LR electrodes ($V_1 \rightarrow C_2$, $V_2\rightarrow C_1$), moves the ion crystal to the left with an equivalent but mirrored trajectory compared to the as-solved rightward shift (Fig.~\ref{fig:fig2}(a)). In this demonstration, the remapping is done in software, though the outcome is identical to if a crossover switch were used. As the horizontal and vertical legs are symmetric but independently wired, there exist equivalent operations for conditional upward or downward shifts that can be implemented in parallel with the conditional horizontal shifts described above. (Fig.~\ref{fig:fig2}(c)).

This C2LR primitive shares the aims and means of the proposal by Malinowski {\it et al} \cite{malinowski2023wire}, using signal broadcasting and dynamic switches to realize favorable scaling of ion transport in a grid trap. The key differences are that the Malinowski proposal is theoretical and suggests maximum flexibility is realized when every electrode has its own switch, whereas the C2LR primitive is shown experimentally to be fully capable of sorting ions in a grid trap using a single effective switch per site.

% A method for scalable ion sorting with similar ambitions was recently proposed by Malinowski {\it et al} \cite{malinowski2023wire} with two key differences from the experimental demonstration here. First merge, swap, and split operations are fundamental primitives in their sorting algorithm, while the proposal here uses only linear transport through junctions. Compared to the 1D merge, swap, and split strategy, these linear shifts--even through junctions \cite{burton2023transport}--are generally faster, more reliable \cite{lancellotti2023low}, and less sensitive to miscalibration and stray fields \cite{ruster2014experimental,murali2020architecting}. Second, the proposal in Ref.~\cite{malinowski2023wire} uses a switchable input on every electrode in a conditional zone, a multi-bit conditional signal, and a separate set of DACs for each condition. Our method, in its full realization, uses one binary signal and one crossover switch per zone with no additional DAC channels. Despite the increased constraints, we show that there is no sacrifice in performance; arbitrary sorting is possible with low motional excitation and high transport speeds.

To demonstrate the performance of the C2LR protocol, we prepare a single Yb-Ba or Ba-Sr crystal in its motional ground state through a combination of Doppler, electromagnetically induced transparency~\cite{lechner2016Electromagnetically} and sideband cooling.  This cooling brings all motional modes to near their quantum ground state, after which we apply C2LR operations in either direction followed by its time reverse to return the ions to the zone center for measurement.  We then perform thermometry using the asymmetry of the motional sidebands to measure the resulting excitation~\cite{Monroe1995Resolved,rasmusson2021optimized}. The measurement ion is $^{171}\mathrm{Yb}^{+}$ for Yb-Ba crystals and $^{137}\mathrm{Ba}^{+}$ for Ba-Sr, and numerical analysis assumes a thermal phonon distribution with mean $\bar{n}$. We focus on the the axial in-phase (XCOM) and axial out-of-phase (XSTR) modes as they are the most relevant modes for current gating schemes \cite{pino2020demonstration}. 

For all presented Yb-Ba measurements, transport sequences begin and end in one of the six horizontal legs where ion fluorescence can be collected with our specific imaging apparatus, while for all Ba-Sr measurements, transport sequences begin and end in the  two central vertical legs where all desired operations can be tested while cooling and measurement are calibrated in the minimal number of locations. These two cases of measurement location, while a result of particularities of the experimental setups, nevertheless demonstrates that the trap itself is capable of high-quality operation in both vertical and horizontal legs. Note that the axial modes of a crystal located in a vertical leg retain the labels XCOM and XSTR despite being rotated in the lab frame.

In Fig.~\ref{fig:fig2}(d), we show the resulting thermal occupation of XCOM and XSTR as a function of speed averaged over all locations and both directions, with this occupation remaining low up to 8~m/s for Ba-Sr and to at least 10~m/s for Yb-Ba. The resulting occupation is dominated by the temperature at which we can prepare the crystals ($\bar{n}_{\mathrm{XCOM}}\approx 0.20$ quanta, $\bar{n}_{\mathrm{XSTR}}\approx 0.03$ quanta), excluding for Ba-Sr XCOM above 8~m/s. In Fig.~\ref{fig:fig2}(e) we show that the occupancy $\bar{n}_\textrm{avg}$ increases linearly as a function of the number of round-trips when running the transport sequence at a fixed speed of 6~m/s, suggesting an incoherent excitation source. The values of the fits are displayed in Table~\ref{TAB:TAB_A}.

All aggregated measurements, including C2LR measurements above and swap or stay measurements below, calculate uncertainties on the average $\bar{n}_\textrm{avg}$ over the $N$ individual measurements $\bar{n}_i$ using one standard deviation on the mean $\sqrt{\mathrm{Var}(n_i)/N}$. These $N$ measurements include variation in shift direction and measurement location.

\section{swap or stay}
\begin{figure}%[!h]
\includegraphics[width =\columnwidth]{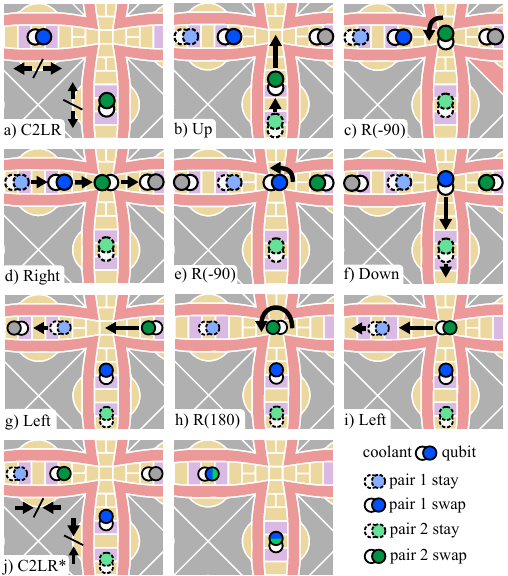}% Here is how to import EPS art
\caption{\label{fig:fig3} Step-by-step depiction of the ion movement during a swap or stay operation.  The pairs are colored based on their initial locations, with saturation indicating whether swap (dark, solid) or stay (light, dashed) was chosen.  All shift and rotation operations are global.  Rotations of the crystal at the junction center are required to maintain the correct crystal order for Yb-Ba but not Ba-Sr crystals. }
\end{figure}

\begin{figure*}%[!h]
\includegraphics[width =\textwidth]{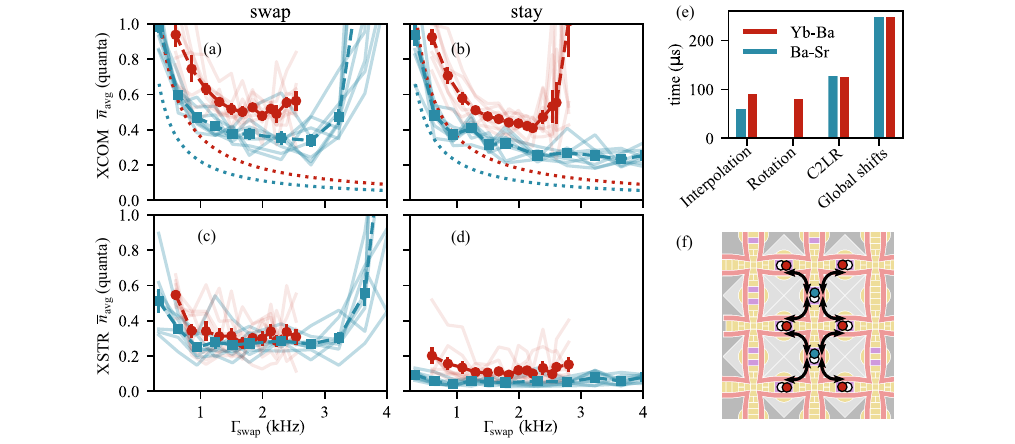}% Here is how to import EPS art
\caption{\label{fig:fig4} Swap or stay performance. (a-d) Resulting occupancy of axial modes following cooling and two swap or two stay operations as a function of swap rate $\Gamma_\mathrm{swap}$. The red (teal) solid lines are sideband asymmetry measurements on a Yb-Ba (Ba-Sr) crystal averaged across all eight swap or stay directions and starting locations ($\bar{n}_\textrm{avg}$), faint lines are the eight individual cases ($\bar{n}_i$), 
and the dotted lines in (a-b) are extrapolated temperatures after an equivalent duration of static heating.  Error bars represent one standard deviation on the spatial average $\bar{n}_\mathrm{avg}$ across all legs to demonstrate the uniformity of the swap or stay operation.  The total swap time ($1/\Gamma_\mathrm{swap}$) is varied by changing the linear transport speed with additional fixed overhead time due to waveform interpolation \cite{burton2023transport} and crystal rotations in the junction. (e) Transport time broken down by operation for Yb-Ba and Ba-Sr at a fixed linear transport speed of 2~m/s.  (f) Despite different starting locations for Yb-Ba (red) and Ba-Sr (teal) crystals, measuring repeated pairs of swap operations means that both setups measure the same eight swap operations (arrows).}
\end{figure*}

We use the C2LR primitive as the key conditional step in a 2D sorting operation that we term `swap or stay'. The swap or stay sequence acts on the entire trap, swapping a target set of diagonally adjacent qubit pairs while returning the others to their initial locations after only intraleg excursions. Swap or stay begins with a C2LR operation and ends with its time reverse. Between the conditional shifts are global transports that primarily transport ions into and out of the junction during which the C2LR electrodes act as generic broadcast electrodes. As these intermediate steps are global, they must be agnostic to the initial C2LR displacement direction and, therefore, accomplish the swap and stay operations in parallel. %If swap is chosen, one half of those wells in each zone is filled with ion crystals, while if stay is chosen the other half is filled.%In short, the C2LR operations select which wells the ions start in, and the global shift then transports these wells to desired destination.

There are four swap or stay variations, each acting on a pair of adjacent crystals but across a different diagonal; for illustration, consider the swap between the crystal that is to the left (pair 1) and the one below (pair 2) the junction (Fig.~\ref{fig:fig3}). The swap begins with a conditional shift toward the junction center for both pairs, i.e. right for pair 1 and up for pair 2, followed by a continued upward shift sending pair 2 into the junction. Next, a rightward shift simultaneously brings pair 2 out of the junction while pair 1 moves in. After this, pair 1 shifts down toward its destination, followed by a leftward shift back through the junction by pair 2. The swap concludes with the reverse of the initial conditional shifts. 

Conversely, the stay operation begins with a conditional shift away from the junction. The global shifts that follow are small enough that neither pair leaves its respective leg. The final shift returns both pairs to their starting locations and, as in the swap case, is the reverse of the initial conditional shift.

In the case of Yb-Ba, there is an additional crystal axis rotation between junction transport operations. This rotation reorients the crystal in the plane of the trap so that the crystal is aligned with the subsequent linear transport direction with Yb facing the junction.  With trajectories and potential well constraints chosen here combined with its larger mass ratio, the Ba-Sr crystals orient normal to the trap plane at the junction center \cite{burton2023transport}, thereby rendering these rotations irrelevant.

Because the intermediate global transport supports either choice of initial C2LR direction, these global shifts  must begin with two wells on each leg of the grid, one at each potential endpoint of the C2LR operation. The C2LR condition chooses which of the wells are occupied after the first step or brought back to the center after the penultimate one. This multiwell transport through junctions in a broadcast grid presents a challenging problem given the demands of multispecies junction transport and the limited number of electrodes per zone. Through judicious prescription of constraints on the potential wells, we successfully generate solutions for the required global shifts with constraints on the wells that enter junctions based upon previous work \cite{burton2023transport}.

We then characterize the motional excitation and achievable speed of the four swap or stay permutations throughout the region of interest. A crystal is moved sequentially to each location, where it is cooled close to its motional ground state. A doubled swap or stay operation at rate $\Gamma_\mathrm{swap}$ is then executed, which unconditionally returns the ions to the measurement zone, and the resulting excitation of the axial motional modes is characterized with sideband asymmetry. The total swap time  $1/\Gamma_\mathrm{swap}$ consists of the active transport duration, interpolation time between steps (10~{\textmu}s per step, Fig.~\ref{fig:fig3}), and, for Yb-Ba, rotation duration (20~{\textmu}s, Fig.~\ref{fig:fig4}(e)). The varied parameter is the transport duration, determined by the average speed and the total effective distance traveled of 750~{\textmu}m, twice the on-grid distance between sites. When testing the doubled swap operations, both crystal types perform the same round-trip operation, differing only in where the round-trip begins (Fig.~\ref{fig:fig4}(f)).

The measured excitation across the swap or stays of interest (Fig.~\ref{fig:fig4}(a-d)) clearly demonstrates that we can perform the operation for both Yb-Ba and Ba-Sr across a range of rates and grid locations while maintaining the crystal near its motional ground state. Furthermore, the spatial uniformity of the performance illustrates the high degree of translation symmetry achieved in the trap. For Yb-Ba, we measure the fastest swap before substantial motional excitation at a rate of 2.5~kHz, corresponding to linear transport speeds of 3.3~m/s, while for Ba-Sr this occurs at 3.2 kHz and a speed of 3.0~m/s. The average resulting excitations are presented in Table~\ref{TAB:TAB_B}. We do not subtract off the substantial contribution of imperfect XCOM ground state cooling (approximately 0.1-0.2 quanta) as the mode occupations mix considerably during junction transits \cite{burton2023transport}. This mixing does imply that measurement of the axial modes serves as an effective gauge of the radial mode excitation.
\begin{table}
\centering
$\begin{array}{@{} S[table-format=3.0] ccccc @{}} 
\toprule
\textrm{Crystal} & \textrm{operation} & \mathrm{XCOM} \ \bar{n}_\textrm{avg}~(\mathrm{q})& \mathrm{XSTR} \ \bar{n}_\textrm{avg}~(\mathrm{q})& 
\\
\midrule

\textrm{Yb-Ba} & \textrm{swap}  & 0.56\pm0.05 & 0.31\pm0.04 \\
\textrm{Yb-Ba} & \textrm{stay}  & 0.53\pm0.09  &  0.10\pm0.02\\ 
\textrm{Ba-Sr} & \textrm{swap}  & 0.48\pm0.04 & 0.30 \pm 0.03\\
\textrm{Ba-Sr} & \textrm{stay}  & 0.25\pm0.06 &  0.08\pm0.03 \\

\bottomrule
\end{array}$
\caption{\label{TAB:TAB_B} Motional excitation in quanta (q) after a doubled swap or stay operation based on sideband asymmetry for Yb-Ba and Ba-Sr crystals at the fastest rate before substantial excitation. The swap or stay excitation at a rate (speed) of 2.5~kHz (3.3~m/s) for Yb-Ba and 3.2~kHz (3.0~m/s) for Ba-Sr} 
\end{table}

Several features worthy of note emerge from these measurements (Fig.~\ref{fig:fig4}(a)-(d)).  For large $\Gamma_\mathrm{swap}$, the excitation sharply increases, a sign of the onset of coherent excitation caused by fast transport \cite{burton2023transport}.  For longer transport times ($\Gamma_\mathrm{swap}<2.0$~kHz), the resulting excitation is dominated by an effective heating rate that is independent of operation. In this regime, the motional excitation scales with total transport time $1/\Gamma_\mathrm{swap}$, particularly for XCOM. This effective heating rate during the transport operations matches closely with the measured static heating rate (Fig.~\ref{fig:fig9}) with an offset from imperfect initial cooling. This coarse agreement demonstrates that the act of transport itself contributes little motional excitation until speeds increase to the onset of coherent excitation.

The XSTR static heating rates are substantially smaller than XCOM; indeed, little heating occurs in XSTR during a stay operation. However, the junction transport present in a swap operation alters the frequency and character of the motional modes, resulting in mixing of motional occupation \cite{burton2023transport} and an observable effective XSTR heating rate.

All data presented in the main text are taken using a single multispecies ion crystal (Yb-Ba or Ba-Sr) that serially samples the numerous wells present in the transport operations and all with identical initial and final measurement locations. As the Coulomb interaction between ions in different wells is small (Appendix~\ref{CoulombInteraction}), such serial testing is an effective proxy for performance of a sort in a densely occupied configuration. Additionally, we validate that the transport operations accomplish their intended outcome in several ways, including a crystal tagging method using Ba-Sr (Appendix~\ref{BaSrTag}). And finally, we visually demonstrate the ability of the swap or stay primitive to sort a fully loaded 2D grid trap in parallel.  We load Yb ions in all eight legs of the region of interest and implement a 2D sorting sequence via swap or stay operations. A stop motion video showing the trajectory of all Yb ions was collected with a complementary metal-oxide-semiconductor (CMOS) camera (see Supplementary Video~S1).

\section{Reorder or not}
In addition to swapping qubits between zones, a trapped ion quantum processor may require zone-specific reordering of ion crystals.  Such conditional reorders may be part of a larger sorting algorithm \cite{kaufmann2017fast}, or simply be used to correct spontaneous crystal reorders that may occur after a background gas collision \cite{hankin2019systematic, moses2023race}. In close analogy with the swap or stay sequence, we demonstrate a reorder or not operation based on C2LR. This operation is demonstrated only for Yb-Ba, though nothing, in principle, precludes extending it to Ba-Sr.

\begin{figure}%[!h]
\includegraphics[width =\columnwidth]{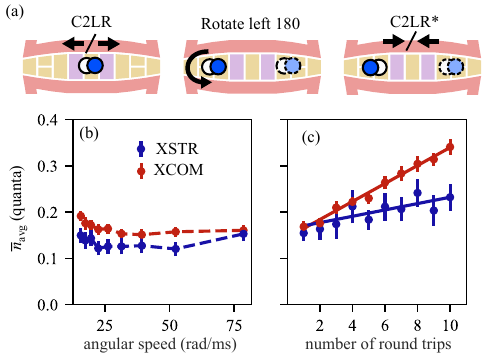}% Here is how to import EPS art
\caption{\label{fig:fig5} Reorder or not description and performance. (a) The sequence begins with C2LR, continues with the a two-well solution where the left well executes a 180 degree rotation while the right well is static, and ends with the reverse of the initial C2LR. (b-c) Average occupancy over measurements zones and conditions of a Yb-Ba crystal following a reorder or not operation as a function of angular speed (b) and number of round-trips (c). The mode occupancy increases approximately linearly with the number of round-trips. Linear fits yield an added noise per reorder of $19.5\pm0.9$~mq/reorder or not for the XCOM mode and $7 \pm 2$~mq/reorder or not for the XSTR mode.  }
\end{figure}
This reorder or not sequence begins and ends with C2LR, as in swap or stay, but with a new intermediate waveform that holds two potential wells but rotates only the left one by $180^\circ$ (Fig.~\ref{fig:fig5}(a)). In Fig.~\ref{fig:fig5}(b) we show the resulting excitation of a Yb-Ba crystal following a reorder or not operation as a function of angular speed, and as a function of number of round-trips in Fig.~\ref{fig:fig5}(c).  In both Fig.~\ref{fig:fig5}(b-c) we average the resulting occupancy over all six horizontal zones of the trap as well as both conditions.  The error bars represent one standard deviation on the average value of the excitation across all measured zones.

From the multiple reorder or not operations performed sequentially, we can infer heating rates per operation and compare this to baseline rates. For C2LR transport speed of 10~m/s and an angular speed of $79$~rad/ms, we infer $(19.5\pm0.9)\times10^{-2}$ quanta/reorder or not in the XCOM mode, and $(7 \pm 2)\times10^{-3}$~quanta/reorder or not for the XSTR mode. Including interpolation time, reorder or not takes 93~{\textmu}s to complete, and with our measured XCOM static heating rate of 181~quanta/s predicts approximately $16\times10^{-3}$~quanta/reorder or not (see Table~\ref{TAB:TAB2} for heating rate data).  The XSTR mode heats less at 12.3~quanta/s, giving approximately $1.1\times10^{-3}$~quanta/reorder or not. As in the case of swap or stay, the heating inferred during transport is close to that of our background heating rates seen in static potential wells with approximately equivalent axial frequencies.    

\section{Conclusion}
Here we have demonstrated the fundamental building blocks for scaling ion transport in a grid-based Paul trap, namely the conditional exchange of crystals between sites and the reorder of crystals within a site, both requiring only a fixed number of analog signals independent of the grid size and one digital input per site. The key novel ingredient in this demonstration is the C2LR primitive, a shift whose direction in a given site can be reversed with the exchange of voltages applied to two particular electrodes. Incorporating C2LR with global transport operations realizes the conditional transport operations presented here, but C2LR could also be combined with location-dependent control fields, such as focused lasers, to realize conditional qubit operations such as initialization, measurement, and gates.  

The success of the transport operations, attaining subquanta motional excitation at 3~m/s speeds across multiple sites in the trap and with crystals of different composition, confirms the necessary control for multispecies junction transport can be realized despite the restrictions imposed by extensive electrode cowiring. Looking forward, scaling to a larger grid should improve transport performance and simplify waveform development, as the trap will more closely resemble a periodic tiling. 

While here the voltages for the different C2LR conditions were exchanged in software, that control can readily be moved in vacuum \cite{alonso2016fast} and even as far downstream as the trap chip itself through the delivery of binary signals and integrated double-pole single-throw switches. Compared to analog signals, the delivery of a large number of binary signals presents minimal technical challenge, and draws directly from application-specific integrated circuit techniques used in classical computing. The number of on-chip switches needed to implement this method for sorting qubits in a large-scale trap are well within the capabilities of cryogenic CMOS technology, and indeed this is already an active area of research for ion traps \cite{stuart2019chip, blain2021hybrid}. 

This demonstration solidifies a key pillar in scaling up the QCCD architecture that, when combined with other crucial scalability efforts such as integrating photonics on chip ~\cite{ivory2021integrated,mehta2016integrated,mordini2024multizone} or controlling qubits without lasers ~\cite{lekitsch2017blueprint,sutherland2024laser}, will bring trapped ion quantum computers from the tens to the thousands of qubits and beyond.

\begin{acknowledgments}
We would like to thank the entire Quantinuum team, in particular contributions from Alex An and Mark Kokish.

\end{acknowledgments}
\bibliography{biblio}% Produces the bibliography via BibTeX.
%\floatbarrier
\appendix
\section{Effective heating rate during transport}
\begin{figure}%[!h]
\includegraphics[width =\columnwidth]{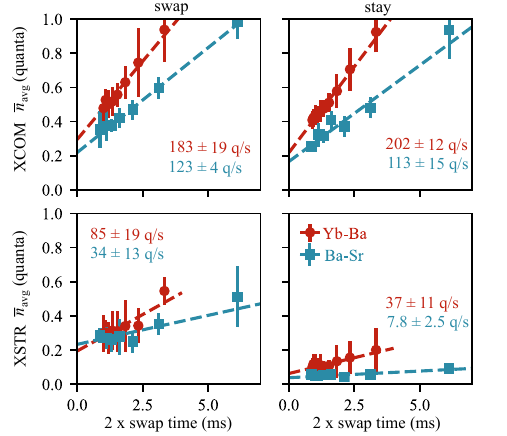}
\caption{\label{fig:fig9} Effective heating rate fits for transport during swap or stay operations for XCOM and XSTR, removing the fastest 4 speeds from the measurements in the main text.  }
\end{figure}

For slow transport speeds ($\Gamma_\mathrm{swap}<2.0$~kHz), we observe a linear increase in excitation with increased transport time that can be described by an effective heating rate.  The linear fits are plotted in Fig.~\ref{fig:fig9} and the resulting effective heating rates are shown in Table~\ref{TAB:TAB2} alongside the measured heating rates in the zone centers for Yb-Ba (Ba-Sr) in multiple horizontal (vertical) zones.  

For both Ba-Sr and Yb-Ba, the effective heating rates of the XCOM modes are comparable to the heating rates measured in a static well at the zone center (Table~\ref{TAB:TAB2}), while XSTR heats significantly more during a swap than a stay. This good XCOM agreement is despite the complicated behavior that occurs as ions approach the junction, involving excursions of the crystal off of the pseudopotential minima, rapidly changing motional mode frequencies and directions, including population exchange between different normal modes.
\begin{table}
\centering
$\begin{array}{@{} S[table-format=3.0] cccccc @{}} 
\toprule
 \mathrm{leg }  & \mathrm{XCOM } & \mathrm{XSTR } & \mathrm{RSTR1 } & \mathrm{RSTR2 } & \mathrm{crystal} \\
\mathrm{(r,c) }  & \mathrm{q/s } & \mathrm{q/s } & \mathrm{q/s } & \mathrm{q/s } \\
\midrule
 \mathrm{(0,0)}  & 141 & 11.8 & 17.9 & 28.2 & \mathrm{Yb-Ba} \\
\mathrm{(0,2)}  & 213  & 12.2 & 37.6 & 28.5 & \mathrm{Yb-Ba}\\
\mathrm{(2,0)} & 169  & 13.3 &  30.6 & 30.0 & \mathrm{Yb-Ba}\\
\mathrm{(2,2)}  & 167 & 10.7 & 27.2 & 31.2 & \mathrm{Yb-Ba}\\
\mathrm{(4,0)}  & 204 & 13.7 & 37.4  & 33.7 & \mathrm{Yb-Ba}\\
\mathrm{(4,2)}  & 191 & 11.9 & 24.4 & 38.4 & \mathrm{Yb-Ba}\\
\mathrm{average}  & 180 \pm 11 &  12.3\pm 0.9 & 29 \pm 3  & 32 \pm 2& \mathrm{Yb-Ba} \\
\mathrm{swap}  & 183 \pm 19  & 85\pm19 & - & -& \mathrm{Yb-Ba} \\
\mathrm{stay}& 202 \pm 12 & 37 \pm 10  & - & - &\mathrm{Yb-Ba} \\
\midrule
 \mathrm{(1,1)} & 91 & 11 & 14 & 35 & \mathrm{Ba-Sr} \\
 \mathrm{(1,3)} & 110  & 8 & 35 & 31 & \mathrm{Ba-Sr}\\
 \mathrm{average} & 101 \pm 27 &  10\pm 3 & 25 \pm 11  & 33 \pm 12 & \mathrm{Ba-Sr}\\
 \mathrm{swap} & 123 \pm 4  & 34 \pm13 & - & - & \mathrm{Ba-Sr}\\
 \mathrm{stay} & 113 \pm 15 & 8 \pm 3  & - & - & \mathrm{Ba-Sr}\\
\bottomrule
\end{array}$
\caption{\label{TAB:TAB2}. Measured heating rates in quanta per second (q/s) across the six horizontal (upper table portion) sites for Yb-Ba and two vertical (lower table portion) sites for Ba-Sr compared to effective heating rates during swap or stay extracted from slow transport operations (Fig.~\ref{fig:fig9})}.   
\end{table}
\section{Modeling ion transport waveforms}
\label{sec:app:waveforms}
As outlined in previous work \cite{burton2023transport}, we generate a numerical electrostatic model that includes the entire trap surface.  This allows us to estimate the total potential as a function of position above the trap: 
\begin{equation}
\phi(\mathbf{x}) = \phi_\mathrm{pp}(\mathbf{x}) + \sum_i \phi_i(\mathbf{x}),
\end{equation}
where $\phi_\mathrm{pp}(\mathbf{x})$ is the pseudopotential resulting from the rf drive and $\phi_i(\mathbf{x})$ is the potential generated by control electrode $i$.  To transport ions between locations in the grid we find multiple potential wells evenly spaced across the transport path through a least-squares numerical optimization procedure.  We then interpolate between these voltages to form waveforms that can be played out on a arbitrary waveform generator to effect ion transport.  

In typical surface electrode traps the ions are constrained to follow paths of minimum pseudopotential to avoid unnecessary micromotion and concomitant heating from noise emitted by the rf source.  However, in junctions it is well studied that the confinement, which is defined as $C = \nabla^2 \phi(\mathbf{x})\propto \sum_i \omega_i$ can drop precipitously \cite{wesenberg2009ideal, burton2023transport} leaving the ion crystal susceptible to noise due to the lowered motional frequencies $\omega_i$.  

It is possible to preserve confinement by allowing the ions to deviate from the paths of minimum pseudopotential \cite{burton2023transport, shu2014heating,wright2013swap}.  This has been used successfully to generate low heating transport waveforms through junction-based ion traps \cite{burton2023transport}, and is the strategy used in this work.
\section{Static potential well frequencies}
A key component of our broadcast electrode scheme is the uniformity/periodicity of the trap zones.  One experimentally accessible proxy for this uniformity, in addition to the success of transport operations in different legs presented in the main text,  is the consistency of the motional mode frequencies from leg to leg (using Yb-Ba).  These frequencies are presented in six legs for Yb-Ba in Table~\ref{TAB:TAB3}. 

\begin{table*}
\centering
$\begin{array}{@{} S[table-format=3.0] ccccc @{}} 
\toprule
\mathrm{leg } & \mathrm{XCOM } & \mathrm{XSTR } & \mathrm{RSTR1 } & \mathrm{RSTR2 } \\
\mathrm{(r,c) } & \mathrm{(MHz)} & \mathrm{(MHz)} & \mathrm{(MHz)} & \mathrm{(MHz)} \\
\midrule
\mathrm{(0,0)} & 1.45944 & 2.56496 &  2.15928 & 2.03187 \\
\mathrm{(0,2)} & 1.46225  & 2.56298 & 2.14316 & 2.02713 \\
\mathrm{(2,0)} & 1.46375   & 2.56526  &  2.17374 & 2.054441 \\
\mathrm{(2,2)} & 1.46440 & 2.56589 & 2.16279  & 2.04457 \\
\mathrm{(4,0)} & 1.46176 & 2.56146& 2.15341   & 2.03401\\
\mathrm{(4,2)} & 1.46060 & 2.55997  & 2.14302 & 2.01260 \\
\mathrm{average} & 1.4620\pm7\times10^{-4}  &  2.5634\pm9\times10^{-4} & 2.155\pm4\times10^{-3}  & 2.034\pm5\times10^{-3}  \\
\bottomrule
\end{array}$
\caption{\label{TAB:TAB3}. The axial center-of-mass and axial/radial stretch mode frequencies of a Yb-Ba crystal in a static well in each leg differ on the single kHz scale.}% (Fig.~\ref{fig:fig9})} 
\end{table*}

\section{Validation of swap or stay transport operations}
\label{BaSrTag}
\subsection{Overview}
Ultimately, grid-based QCCD devices will need to be run with all horizontal and vertical legs filled with ions, but a discussion of the required sorting algorithms and hardware for such a device is beyond the scope of this work.  All transport operations presented in the main text are demonstrated using a single Yb-Ba or Ba-Sr crystal, and here we discuss various validation experiments to confirm that the ions are undergoing the correct swap or stay transport operations.  

It first should be pointed out that we can run transport with the Yb-Ba (Ba-Sr) crystals remaining in approximately their quantum ground state following a swap or stay operation, and that this agrees well with our numerical simulations.  This already provides strong evidence that the swap or stay is being completed as expected, but we demonstrate additional validation steps below.  

\subsection{Parallel Ba-Sr ion crystal swapping}
To directly verify the intended outcome of the operation in parallel operation, two Ba-Sr crystals were load into adjacent horizontal and vertical zones. The two crystals are Doppler cooled using a pair of copropagating 422~nm and 493~nm beams aligned at 45 degrees to the grid axis, illuminating both crystals simultaneously, with repump beams (at 1092~nm and 650~nm) delivered as sheet beams. A single 1762~nm beam aligned with the vertical leg drives transitions between the $^6S_{1/2}$ to $^5D_{5/2}$ states of the Ba ion. The $^5D_{5/2}$ state is isolated from the fluorescence cycling transitions, and so the ion is dark if driven to this 'shelved' state \cite{an2022high}. 

\begin{figure}[!h]
\includegraphics[width = \columnwidth]{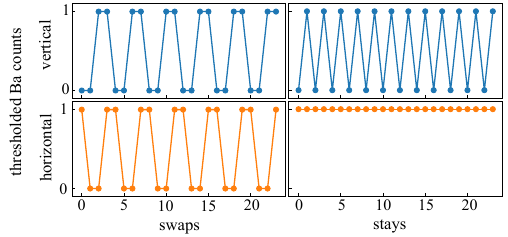}
\caption{\label{fig:fig11} The Ba fluorescence of two Ba-Sr crystals in two diagonally adjacent legs is measured after a number of either swap or stay operations. The crystals are `tagged'  using $1762$ nm light resonant with the shelving transition $^6S_{1/2}$ to $^5D_{5/2}$. A $\pi$-pulse on this transition is driven just before fluorescence is measured. During the repeated stay operations, this results in the Ba in the vertical leg oscillating between dark (shelved) and bright (ground state) with a period of two, while both legs oscillate with twice the period during the repeated swap operation}
\end{figure}

A sequence of either swaps or stays with a $\pi$ pulse on the shelving transition results in a clear demonstration of successful swaps and stays with two crystals in parallel. During repeated stay operations, the Ba in the vertical leg is driven to the $^5D_{5/2}$ state or back every cycle, and therefore oscillates between bright or dark with a period of two operations, while the Ba in the horizontal leg remains bright at all times. For the swap case, the Ba that is initially shelved in the vertical zone then swaps to the horizontal zone, where it remains dark until it returns to the vertical zone after the next swap (Fig.~\ref{fig:fig11}). The toggle between bright and dark is half as fast because it was twice the ions to shelve, resulting in oscillations between bright and dark in both zones with twice the period (4 operations) of the stay case.

\subsection{Single Yb transport interrogated by a Doppler sheet beam}
A third experimental setup and nominally identical trap goes farther and demonstrates basic sorting using the swap or stay operation. A 369~nm sheet beam is used to address single $^{171}\mathrm{Yb}$ ions in a fully loaded section of a $3\times 2$ section of the grid trap, and the resulting fluorescence was detected by a CMOS camera spanning the entire grid section of the trap.  In Supplementary Video~S1, using the exact same voltage waveforms for the Yb-Ba data shown in the main text, we show swapping of Yb ions between the vertical zones and horizontal zones above and to the left.  We choose to record the video using a third experimental setup as this system had a sheet beam and camera setup that was capable of imaging the entire $3\times2$ grid section of the trap.  Single Yb ions are used since a Ba sheet beam was not immediately available.  

\subsection{Stray fields from other ion crystals}\label{CoulombInteraction}
As shown in Fig.~\ref{fig:fig3}, during a swap or stay operation the ions are continually shuttled into and out of the junction during the swap procedure.  This leaves open the possibility of Coulomb interactions between neighbouring ion crystals affecting transport.  To bound the effects of this interaction we can use the minimum separation between two ion crystals, which occurs when one crystal is in the junction and the other is in a static horizontal or vertical well--see for example Fig.~\ref{fig:fig3}, step (e).  In this configuration the separation between the two crystals is $d=$\SI{125}{\micro\metre} and stray electric field then consists of the direct Coulomb interaction, and the Coulomb field resulting from the effective image charge generated by the trap top metal  
\begin{equation}
\mathbf{E}_\textrm{stray} = \frac{2e}{4\pi\epsilon_{o}}\left(\left(\frac{1}{d^2} - \frac{d}{(d^2 + 4h^2)^{3/2}}\right)\hat{\mathbf{x}}
 -\frac{2h}{(d^2 + 4h^2)^{3/2}}\hat{\mathbf{z}}\right).
\end{equation}

For $d=$~\SI{125}{\micro\metre} and $h=$~\SI{53}{\micro\metre} the magnitude of the stray field is $|E_\mathrm{stray}|=0.12$~V/m, where $e$ is the electron charge.  As stray fields from surface imperfections, local charging and other effects are typically in the range of 10-100~V/m in surface-electrode traps \cite{doret2012controlling}, the effective stray field due to Coulomb potentials can safely be ignored, even in a densely loaded grid.  This indicates that measurements of transport of a single multispecies ion crystal throughout the trap faithfully represents the motional heating one would expect to measure in a fully loaded trap.

\end{document}